**Version 1.0**

VLADIMIR V. KOROSTELEV

# A Primer in Quantum Mechanics for NMR Students

# Table of Contents

## 1. Introduction

In classical physics, the physical states of an object of interest can be defined exactly to a degree, which is mainly limited by experimental factors such as random and systematic errors. The measurement of a physical state in quantum mechanics is different as it includes intrinsic uncertainty, which cannot be influenced by improvements in experimental techniques (*1*). This concept, originally proposed by Heisenberg (*2*), is called the Uncertainty Principle, which states that the uncertainties of measurement of energy $\Delta E$ and interval of time $\Delta t$, during which a microscopic particle possesses that energy, relate as:

$$\Delta E \cdot \Delta t \geq h \tag{0}$$

where *h* is Planck's constant. Hence, only the probabilities of getting particular results can be obtained in quantum mechanical experiments, fundamentally distinguishing them from the classical ones. The uncertainties in quantum-mechanical measurements stem from the disturbance which measurement itself causes to the measured state at a microscopic scale. In NMR, the life times of spin states do not generally exceed the spin-lattice relaxation time $T_1$, and therefore the half-widths of NMR lines in spectra must be at least of the order of $1/T_1$ (*3*).

The uncertainties featured in quantum-mechanical measurements lead to a probability interpretation of phenomena, where the quantum-mechanical states are described by wave functions given in particular representation. In Dirac formulism of quantum mechanics (*4*) used throughout this text, a state of a quantum mechanical system is described by vector called ket and written as $|\psi\rangle$. The use of ket instead of wave function allows the form of analysis, which is independent of the particular representation chosen. In this formulism, different representations are regarded as rotations in vector space, hence the ket $|\psi\rangle$ represents quantum state no matter what representation is chosen for the analysis.

The number of the ket components is $2 \cdot I + 1$, where *I* is spin. Thus, for spin ½ ket has two components, each of which is a complex number[1].

## 2. Quantum states of a spin ½

Single spin ½ in static magnetic field acting along *z*-axis has two eigenkets: $|\alpha\rangle$ (spin along the field) and $|\beta\rangle$ (spin is opposite to the field) described by columns:

$$|\alpha\rangle = \begin{pmatrix} 1 \\ 0 \end{pmatrix} \tag{1}$$

---

[1] when imaginary part is zero the number is represented by its real part only.

$$|\beta\rangle = \begin{pmatrix} 0 \\ 1 \end{pmatrix} \qquad (2)$$

In general, the state of spin ½ may be represented by combination of eigenkets written as the ket $|\psi\rangle$ where

$$|\psi\rangle = c_1|\alpha\rangle + c_2|\beta\rangle = c_1\begin{pmatrix}1\\0\end{pmatrix} + c_2\begin{pmatrix}0\\1\end{pmatrix} = \begin{pmatrix}c_1\\0\end{pmatrix} + \begin{pmatrix}0\\c_2\end{pmatrix} = \begin{pmatrix}c_1\\c_2\end{pmatrix} \qquad (3),$$

$c_1$ and $c_2$ are complex numbers which relate to probabilities of a spin to be in the particular state, either in state $|\alpha\rangle$ or in state $|\beta\rangle$. The eigenkets are special kets, which are normally orthogonal and represent the states in which a quantum-mechanical system can be found when its state is measured. The particular eigenkets can be chosen for description of a quantum mechanical state, like a frame of reference. However, any chosen set of eigenkets must be normalised and complete, in order to be appropriate for a representation of a quantum mechanical state. (The conditions of orthogonality, normalization and completeness of kets will be presented here.)

A bra vector, written as $\langle\psi|$ is defined in separate space of bras $\langle\alpha|$ and $\langle\beta|$ where

$$\langle\alpha| = \begin{pmatrix}1 & 0\end{pmatrix} \qquad (4)$$

$$\langle\beta| = \begin{pmatrix}0 & 1\end{pmatrix} \qquad (5)$$

and may be represented as:

$$\langle\psi| = c_1^*\langle\alpha| + c_2^*\langle\beta| = c_1^*\begin{pmatrix}1 & 0\end{pmatrix} + c_2^*\begin{pmatrix}0 & 1\end{pmatrix} = \begin{pmatrix}c_1^* & 0\end{pmatrix} + \begin{pmatrix}0 & c_2^*\end{pmatrix} = \begin{pmatrix}c_1^* & c_2^*\end{pmatrix}$$

(6)

Thus, two scalar[2] products can be formed (here notation of $\langle\psi|\cdot|\psi\rangle$ and $|\psi\rangle\cdot\langle\psi|$ is shorten to $\langle\psi|\psi\rangle$ and $|\psi\rangle\langle\psi|$ respectively[3]):

---

[2] scalar products are used for spin states of non-interacting spins. For states of coupled spins tensor products are used.
[3] superscript stars denote complex conjugation

$$\langle\psi|\psi\rangle = \begin{pmatrix} c_1^* & c_2^* \end{pmatrix} \cdot \begin{pmatrix} c_1 \\ c_2 \end{pmatrix} = c_1^* \cdot c_1 + c_2^* \cdot c_2 \tag{7}$$

For example:

$$\langle\alpha|\alpha\rangle = \begin{pmatrix} 1 & 0 \end{pmatrix} \cdot \begin{pmatrix} 1 \\ 0 \end{pmatrix} = 1 \cdot 1 + 0 \cdot 0 = 1 \tag{8}$$

$$\langle\alpha|\beta\rangle = \begin{pmatrix} 1 & 0 \end{pmatrix} \cdot \begin{pmatrix} 0 \\ 1 \end{pmatrix} = 1 \cdot 0 + 0 \cdot 1 = 0 \tag{9}$$

$$\langle\beta|\alpha\rangle = \begin{pmatrix} 0 & 1 \end{pmatrix} \cdot \begin{pmatrix} 1 \\ 0 \end{pmatrix} = 0 \cdot 1 + 1 \cdot 0 = 0 \tag{10}$$

$$\langle\beta|\beta\rangle = \begin{pmatrix} 0 & 1 \end{pmatrix} \cdot \begin{pmatrix} 0 \\ 1 \end{pmatrix} = 0 \cdot 0 + 1 \cdot 1 = 1 \tag{11}$$

Also a matrix of an operator (more on operators will be explained in next section) can be formed:

$$|\psi\rangle\langle\psi| = \begin{pmatrix} c_1 \\ c_2 \end{pmatrix} \cdot \begin{pmatrix} c_1^* & c_2^* \end{pmatrix} = \begin{pmatrix} c_1 \cdot c_1^* & c_1 \cdot c_2^* \\ c_2 \cdot c_1^* & c_2 \cdot c_2^* \end{pmatrix} \tag{12}$$

For example:

$$|\alpha\rangle\langle\alpha| = \begin{pmatrix} 1 \\ 0 \end{pmatrix} \cdot \begin{pmatrix} 1 & 0 \end{pmatrix} = \begin{pmatrix} 1 \cdot 1 & 1 \cdot 0 \\ 0 \cdot 1 & 0 \cdot 0 \end{pmatrix} = \begin{pmatrix} 1 & 0 \\ 0 & 0 \end{pmatrix} \tag{13}$$

$$|\alpha\rangle\langle\beta| = \begin{pmatrix} 1 \\ 0 \end{pmatrix} \cdot \begin{pmatrix} 0 & 1 \end{pmatrix} = \begin{pmatrix} 1 \cdot 0 & 1 \cdot 1 \\ 0 \cdot 0 & 0 \cdot 1 \end{pmatrix} = \begin{pmatrix} 0 & 1 \\ 0 & 0 \end{pmatrix} \tag{14}$$

$$|\beta\rangle\langle\alpha| = \begin{pmatrix} 0 \\ 1 \end{pmatrix} \cdot \begin{pmatrix} 1 & 0 \end{pmatrix} = \begin{pmatrix} 0 \cdot 1 & 0 \cdot 0 \\ 1 \cdot 1 & 1 \cdot 0 \end{pmatrix} = \begin{pmatrix} 0 & 0 \\ 1 & 0 \end{pmatrix} \tag{15}$$

$$|\beta\rangle\langle\beta| = \begin{pmatrix} 0 \\ 1 \end{pmatrix} \cdot \begin{pmatrix} 0 & 1 \end{pmatrix} = \begin{pmatrix} 0 \cdot 0 & 0 \cdot 1 \\ 1 \cdot 0 & 1 \cdot 1 \end{pmatrix} = \begin{pmatrix} 0 & 0 \\ 0 & 1 \end{pmatrix} \tag{16}$$

Products (13) and (16) are matrix representations of operators called *projection operators* and denoted as $\hat{I}^\alpha$ and $\hat{I}^\beta$, respectively, i.e.

$$\hat{I}^\alpha = |\alpha\rangle\langle\alpha| \quad \text{and} \quad \hat{I}^\beta = |\beta\rangle\langle\beta| \tag{17-18}.$$

Also, products (14) and (15) are matrix representations of operators called *shift operators* denoted as $\hat{I}^+$ and $\hat{I}^-$, respectively, i.e.

$$\hat{I}^+ = |\alpha\rangle\langle\beta| \quad \text{and} \quad \hat{I}^- = |\beta\rangle\langle\alpha| \tag{19-20}.$$

Operator $|\phi\rangle\langle\varphi|$ (in similar with (12)) may act on a bra from the right to form a complex number times the bra $\langle\varphi|$, i.e.

$$\langle\psi|\phi\rangle\langle\varphi| = g_1\langle\varphi| \tag{21}$$

where $g_1$ is a complex number. For example, for $\langle\psi| = \begin{pmatrix} k_1^* & k_2^* \end{pmatrix}$ and $|\phi\rangle = \begin{pmatrix} f_1 \\ f_2 \end{pmatrix}$ the equation (21) determines complex number $g_1$ as $g_1 = k_1^* \cdot f_1 + k_2^* \cdot f_2$, (where $k_1^*$, $k_2^*$, $f_1$, $f_2$ are complex numbers).

Or $|\phi\rangle\langle\varphi|$ can act on a ket from the left to form a complex number times the ket $|\phi\rangle$, i.e.

$$|\phi\rangle\langle\varphi|\psi\rangle = g_2|\phi\rangle \tag{22},$$

where for $\langle\varphi| = \begin{pmatrix} n_1^* & n_2^* \end{pmatrix}$ and $|\psi\rangle = \begin{pmatrix} r_1 \\ r_2 \end{pmatrix}$, the complex number $g_2 = n_1^* \cdot r_1 + n_2^* \cdot r_2$ (where $n_1^*$, $n_2^*$, $r_1$, $r_2$ are complex numbers) according to the rule given by (7).

Scalar product of state with itself (7) always yields a real number and the positive square root of this number is called the norm of the state vector. When state is normalized its norm is unity, i.e.

$$\sqrt{\langle\psi|\psi\rangle} = 1 \tag{23}$$

States $|\alpha\rangle$ and $|\beta\rangle$ are normalized as their norms, $\langle\alpha|\alpha\rangle$ and $\langle\beta|\beta\rangle$, respectively are equal to unity as it is shown in (8) and (11).

States are orthogonal when their product gives zero; for example, the states $|\alpha\rangle$ and $|\beta\rangle$ are orthogonal as can be seen from (7) and (8).

When a quantum mechanical system (a spin, for example) is certainly present in a set of eigenkets, the set is called complete and the sum of diagonal elements of matrix, given by products of bras and kets is unity. This sum represents the total probability of finding a particle in the basis states. Hence, if set of kets $|\alpha\rangle$ and $|\beta\rangle$ is complete then state $|\psi\rangle$ given by (3) is complete as well and sum of diagonal elements of matrix given by (10) has to be unity, i.e.

$$c_1 \cdot c_1^* + c_2 \cdot c_2^* = 1 \qquad (24)$$

## 3. Operators for spin ½

A new concept, specific to quantum mechanics, relates to the way of calculating the values of dynamical variables measured in experiment. In classical physics, the dynamic variables are normally described by functions, as opposed to quantum mechanics, where dynamical variables are represented in terms of expectation values of their operators.

An operator is a symbolic representation of a mathematical operation accomplished on one or more functions. Only linear operators are used in quantum mechanics and only Hermitian operators (they are linear and equal to their adjoints) to represent dynamical variables. In the matrix version of quantum mechanics, operators are represented by square matrices, whose elements correspond to the different states of quantum mechanical system.

When linear operator is applied to a ket or bra it results in a new ket or bra, respectively. This can be written for operator $\hat{A}$ and kets $|\psi_1\rangle$ and $|\psi_2\rangle$ in the form:

$$\hat{A}|\psi_1\rangle = |\psi_2\rangle \qquad (25)$$

where $|\psi_1\rangle$ and $|\psi_2\rangle$ are the kets, which represent the original and the new quantum-mechanical states, respectively. For example, if operator $\hat{A}$ is given by matrix $A$:

$$A = \begin{pmatrix} a_{11} & a_{12} \\ a_{21} & a_{22} \end{pmatrix} \qquad (26)$$

and $|\psi_1\rangle$ by column vector:

$$|\psi_1\rangle = \begin{pmatrix} c_1 \\ c_2 \end{pmatrix} \qquad (27)$$

then

$$|\psi_2\rangle = A \cdot |\psi_1\rangle = \begin{pmatrix} a_{11} & a_{12} \\ a_{21} & a_{22} \end{pmatrix} \cdot \begin{pmatrix} c_1 \\ c_2 \end{pmatrix} = \begin{pmatrix} a_{11} \cdot c_1 + a_{12} \cdot c_2 \\ a_{21} \cdot c_1 + a_{22} \cdot c_2 \end{pmatrix} \qquad (28)$$

In summary, linear operators are used for the quantum-mechanical description of physical interactions and dynamical variables, while quantum-mechanical states are represented by wave functions or kets, in Dirac formulism used here. A result of the measurement of a dynamical variable is one of the eigenvalues of the operator, which represents this dynamical variable. The eigenvalues of a linear operator $\hat{A}$ can be found solving its eigenvalue equation:

$$\hat{A}|\psi\rangle = A_n|\psi\rangle \qquad (29)$$

which links the known operator $\hat{A}$ and two unknowns: its eigenvalues and their corresponding eigenkets.

For example, the spin operators $\hat{I}_x$, $\hat{I}_y$, $\hat{I}_z$ for spin ½ are represented by spin matrices:

$$I_x = \begin{pmatrix} 0 & \tfrac{1}{2} \\ \tfrac{1}{2} & 0 \end{pmatrix} \qquad (30)$$

$$I_y = \begin{pmatrix} 0 & -i/2 \\ i/2 & 0 \end{pmatrix} \qquad (31)$$

$$I_z = \begin{pmatrix} \tfrac{1}{2} & 0 \\ 0 & -\tfrac{1}{2} \end{pmatrix} \qquad (32)$$

Hence, we can calculate how these operator change $|\alpha\rangle$ and $|\beta\rangle$ states, for example, using (26):

$$\hat{I}_x|\alpha\rangle = \begin{pmatrix} 0 & \frac{1}{2} \\ \frac{1}{2} & 0 \end{pmatrix} \cdot \begin{pmatrix} 1 \\ 0 \end{pmatrix} = \begin{pmatrix} 0 \cdot 1 + \frac{1}{2} \cdot 0 \\ \frac{1}{2} \cdot 1 + 0 \cdot 0 \end{pmatrix} = \begin{pmatrix} 0 \\ \frac{1}{2} \end{pmatrix} = \frac{1}{2} \cdot \begin{pmatrix} 0 \\ 1 \end{pmatrix} = \frac{1}{2}|\beta\rangle \tag{33}$$

$$\hat{I}_x|\beta\rangle = \begin{pmatrix} 0 & \frac{1}{2} \\ \frac{1}{2} & 0 \end{pmatrix} \cdot \begin{pmatrix} 0 \\ 1 \end{pmatrix} = \begin{pmatrix} 0 \cdot 0 + \frac{1}{2} \cdot 1 \\ \frac{1}{2} \cdot 0 + 0 \cdot 1 \end{pmatrix} = \begin{pmatrix} \frac{1}{2} \\ 0 \end{pmatrix} = \frac{1}{2} \cdot \begin{pmatrix} 1 \\ 0 \end{pmatrix} = \frac{1}{2}|\alpha\rangle \tag{34}$$

$$\hat{I}_y|\alpha\rangle = \begin{pmatrix} 0 & -\frac{i}{2} \\ \frac{i}{2} & 0 \end{pmatrix} \cdot \begin{pmatrix} 1 \\ 0 \end{pmatrix} = \begin{pmatrix} 0 \cdot 1 - \frac{i}{2} \cdot 0 \\ \frac{i}{2} \cdot 1 + 0 \cdot 0 \end{pmatrix} = \begin{pmatrix} 0 \\ \frac{i}{2} \end{pmatrix} = \frac{i}{2} \cdot \begin{pmatrix} 0 \\ 1 \end{pmatrix} = \frac{i}{2}|\beta\rangle \tag{35}$$

$$\hat{I}_y|\beta\rangle = \begin{pmatrix} 0 & -\frac{i}{2} \\ \frac{i}{2} & 0 \end{pmatrix} \cdot \begin{pmatrix} 0 \\ 1 \end{pmatrix} = \begin{pmatrix} 0 \cdot 0 - \frac{i}{2} \cdot 1 \\ \frac{i}{2} \cdot 0 + 0 \cdot 1 \end{pmatrix} = \begin{pmatrix} -\frac{i}{2} \\ 0 \end{pmatrix} = -\frac{i}{2} \cdot \begin{pmatrix} 1 \\ 0 \end{pmatrix} = -\frac{i}{2}|\alpha\rangle \tag{36}$$

$$\hat{I}_z|\alpha\rangle = \begin{pmatrix} \frac{1}{2} & 0 \\ 0 & -\frac{1}{2} \end{pmatrix} \cdot \begin{pmatrix} 1 \\ 0 \end{pmatrix} = \begin{pmatrix} \frac{1}{2} \cdot 1 + 0 \cdot 0 \\ 0 \cdot 1 - \frac{1}{2} \cdot 0 \end{pmatrix} = \begin{pmatrix} \frac{1}{2} \\ 0 \end{pmatrix} = \frac{1}{2} \cdot \begin{pmatrix} 1 \\ 0 \end{pmatrix} = \frac{1}{2}|\alpha\rangle \tag{37}$$

$$\hat{I}_z|\beta\rangle = \begin{pmatrix} \frac{1}{2} & 0 \\ 0 & -\frac{1}{2} \end{pmatrix} \cdot \begin{pmatrix} 0 \\ 1 \end{pmatrix} = \begin{pmatrix} \frac{1}{2} \cdot 0 + 0 \cdot 1 \\ 0 \cdot 0 - \frac{1}{2} \cdot 1 \end{pmatrix} = \begin{pmatrix} 0 \\ -\frac{1}{2} \end{pmatrix} = -\frac{1}{2} \cdot \begin{pmatrix} 0 \\ 1 \end{pmatrix} = -\frac{1}{2}|\beta\rangle \tag{38}$$

## 4. Hamiltonian of a spin in the static magnetic field

Hamiltonian operator represents energy of a quantum mechanical system and is used for description of the physical interactions and time evolution of the system. Its eigenvalues are found by solving the time independent Schrödinger equation, given by:

$$\hat{H}|\psi\rangle = E_n|\psi\rangle \tag{39}$$

The unknowns in (2.15) are the eigenvalues $E_n$ and the eigenfunctions $|\psi\rangle$. When a time-independent Hamiltonian is applied to a ket $|\psi\rangle$, the latter changes according to the time-dependent Schrödinger equation:

$$i\hbar \frac{\partial|\psi(t)\rangle}{\partial t} = \hat{H}|\psi(t)\rangle \tag{40}.$$

When ket is either an eigenket of the Hamiltonian or linear combination of such eigenkets, the solution is:

$$|\psi(t)\rangle = U(t)|\psi(0)\rangle \quad (41),$$

where $|\psi(0)\rangle$ is the ket at $t=0$, and

$$U(t) = e^{-t\frac{\hat{H}t}{\hbar}} \quad (42)$$

is the propagator, which describes the time evolution of a quantum mechanical system.

When Hamiltonian $\hat{H}$ is given in $\hbar$ units, like (30-32), then (42) simplifies to:

$$U(t) = e^{-i\hat{H}t} \quad (43).$$

Time evolution of ket is expressed as change of vector ket's componets. When ket does not change with time its components are complex numbers (sometimes – real numbers, if imaginary part of complex number is zero) and if ket does change with time then its components become functions of time. Propagators are linear operators, and generally can be non-Hermitian.

For a spin placed in static magnetic field acting along *z*-axis, the Hamiltonian operator is given by (*5*):

$$\hat{H}_Z = \omega_{0L} \cdot \hat{I}_z \quad (44)$$

where

$$\omega_{0L} = -\gamma \cdot B_0 \quad (45)$$

is the Larmor frequency, which is the frequency of spin precession in the static magnetic field and $B_0$ represents magnitude of the static magnetic field.

Besides external static magnetic field $B_0$ spins in molecules experience local fields, which alter Larmor frequency and lead to chemically shifted Larmor frequency (*6*). Although this effect is crucial for NMR, it is beyond of scope of this text and hence will be neglected.

**Example:** Find Larmor frequency of proton in the static magnetic field 0.2 T.

*Solution:*
$$\omega_L = -\gamma \cdot |\vec{B}_0|$$
$$\gamma_p = 267.522 \cdot 10^6 \left[\frac{rad}{s \cdot T}\right]$$
thus,
$$\omega_L = -267.522 \cdot 0.2 \cdot 10^6 = -53.5044 \cdot 10^6 \left[\frac{rad}{s}\right],$$
in Hz: $\nu_L = \frac{|\omega_L|}{2 \cdot \pi} = \left|-\frac{53.5044}{2 \cdot \pi} \cdot 10^6\right| \cong 8.5 \cdot 10^6 [Hz] = 8.5 [MHz].$

For spin ½ matrix of Hamiltonian operator given by (44) is:

$$H_Z = \begin{pmatrix} \omega_{0L}/2 & 0 \\ 0 & -\omega_{0L}/2 \end{pmatrix} \qquad (46)$$

As matrix of this Hamiltonian is diagonal the matrix of corresponding propagator is diagonal too:

$$U(t) = \begin{pmatrix} e^{-\frac{i\omega_{0L}}{2}} & 0 \\ 0 & e^{\frac{i\omega_{0L}}{2}} \end{pmatrix} \qquad (47)$$

In general, when matrix of Hamiltonian is non-diagonal the matrix of its propagator given by either (42) or (43) is non-diagonal as well.

For arbitrary ket given by (3) the solution of time-dependent Schrödinger equation (given by (41)) can be found as a product of propagator matrix and ket (3):

$$|\psi(t)\rangle = U(t) \cdot |\psi(0)\rangle = \begin{pmatrix} e^{-\frac{i\omega_{0L}t}{2}} & 0 \\ 0 & e^{\frac{i\omega_{0L}t}{2}} \end{pmatrix} \cdot \begin{pmatrix} c_1 \\ c_2 \end{pmatrix} = \\ \begin{pmatrix} c_1 \cdot e^{-\frac{i\omega_{0L}t}{2}} + 0 \cdot c_2 \\ 0 \cdot c_1 + c_2 \cdot e^{\frac{i\omega_{0L}t}{2}} \end{pmatrix} = \begin{pmatrix} c_1 \cdot e^{-\frac{i\omega_{0L}t}{2}} \\ c_2 \cdot e^{\frac{i\omega_{0L}t}{2}} \end{pmatrix} \qquad (48)$$

This demonstrates that as a result of evolution the components of ket become time-dependent through exponentials being multiplied on components of initial ket.

## 5. Observables and density matrix

Quantum-mechanical states and operators relate to experimental results through observables. These are defined as expectation values of operators of dynamical variables whose values are measured in experiments. In quantum mechanics, "any result of a measurement of a real dynamical variable is one of its eigenvalues. Conversely, every eigenvalue is a possible result of a measurement of the dynamical variable for some state of the system"(*7*). The expectation value of the observable for an operator $\hat{A}$ can be represented in the form:

$$\langle \hat{A} \rangle = \langle \psi | \hat{A} | \psi \rangle \tag{49}$$

Thus, the expectation value of operator $\hat{A}$ given by matrix

$$A = \begin{pmatrix} a_{11} & a_{12} \\ a_{21} & a_{22} \end{pmatrix} \tag{50}$$

can be computed for the case where ket is $|\psi\rangle = \begin{pmatrix} c_1 \\ c_2 \end{pmatrix}$ as follows:

$$\langle \hat{A} \rangle = \langle \psi | \hat{A} | \psi \rangle = \begin{pmatrix} c_1^* & c_2^* \end{pmatrix} \cdot \begin{pmatrix} a_{11} & a_{12} \\ a_{21} & a_{22} \end{pmatrix} \cdot \begin{pmatrix} c_1 \\ c_2 \end{pmatrix} =$$
$$\begin{pmatrix} c_1^* \cdot a_{11} + c_2^* \cdot a_{21} & c_1^* \cdot a_{12} + c_2^* \cdot a_{22} \end{pmatrix} \cdot \begin{pmatrix} c_1 \\ c_2 \end{pmatrix} = \tag{51}$$
$$c_1 \cdot c_1^* \cdot a_{11} + c_1 \cdot c_2^* \cdot a_{21} + c_1^* \cdot c_2 \cdot a_{12} + c_2 \cdot c_2^* \cdot a_{22}$$

A good way of seeing how this works is to do a particular example. Suppose that we want to calculate the expectation value of opearator $\hat{I}_z$ given by

$$\hat{I}_z = \begin{pmatrix} \tfrac{1}{2} & 0 \\ 0 & -\tfrac{1}{2} \end{pmatrix} \tag{52}$$

where $a_{11}=1/2$, $a_{12}=0$, $a_{21}=0$ and $a_{22}=-1/2$. Then

$$\langle \hat{I}_z \rangle = \langle \psi | \hat{I}_z | \psi \rangle = \begin{pmatrix} c_1^* & c_2^* \end{pmatrix} \cdot \begin{pmatrix} \tfrac{1}{2} & 0 \\ 0 & -\tfrac{1}{2} \end{pmatrix} \cdot \begin{pmatrix} c_1 \\ c_2 \end{pmatrix} =$$
$$\begin{pmatrix} c_1^* \cdot \tfrac{1}{2} + c_2^* \cdot 0 & c_1^* \cdot 0 + c_2^* \cdot \left(-\tfrac{1}{2}\right) \end{pmatrix} \cdot \begin{pmatrix} c_1 \\ c_2 \end{pmatrix} =$$
$$c_1 \cdot c_1^* \cdot \tfrac{1}{2} + c_1 \cdot c_2^* \cdot 0 + c_1^* \cdot c_2 \cdot 0 + c_2 \cdot c_2^* \cdot \left(-\tfrac{1}{2}\right) =$$
$$\tfrac{1}{2} \cdot c_1 \cdot c_1^* - \tfrac{1}{2} \cdot c_2 \cdot c_2^*$$
(53).

Using (51) we can also work out the expectation values of operators $\hat{I}_x$ and $\hat{I}_y$

$$\langle \hat{I}_x \rangle = \langle \psi | \hat{I}_x | \psi \rangle = \begin{pmatrix} c_1^* & c_2^* \end{pmatrix} \cdot \begin{pmatrix} 0 & \tfrac{1}{2} \\ \tfrac{1}{2} & 0 \end{pmatrix} \cdot \begin{pmatrix} c_1 \\ c_2 \end{pmatrix} =$$
$$\begin{pmatrix} c_1^* \cdot 0 + c_2^* \cdot \tfrac{1}{2} & c_1^* \cdot \tfrac{1}{2} + c_2^* \cdot 0 \end{pmatrix} \cdot \begin{pmatrix} c_1 \\ c_2 \end{pmatrix} =$$
$$c_1 \cdot c_1^* \cdot 0 + c_1 \cdot c_2^* \cdot \tfrac{1}{2} + c_1^* \cdot c_2 \cdot \tfrac{1}{2} + c_2 \cdot c_2^* \cdot 0 =$$
$$\tfrac{1}{2} \cdot c_1 \cdot c_2^* + \tfrac{1}{2} \cdot c_1^* \cdot c_2$$
(54),

$$\langle \hat{I}_y \rangle = \langle \psi | \hat{I}_y | \psi \rangle = \begin{pmatrix} c_1^* & c_2^* \end{pmatrix} \cdot \begin{pmatrix} 0 & -\tfrac{i}{2} \\ \tfrac{i}{2} & 0 \end{pmatrix} \cdot \begin{pmatrix} c_1 \\ c_2 \end{pmatrix} =$$
$$\begin{pmatrix} c_1^* \cdot 0 + c_2^* \cdot \tfrac{i}{2} & c_1^* \cdot \left(\tfrac{-i}{2}\right) + c_2^* \cdot 0 \end{pmatrix} \cdot \begin{pmatrix} c_1 \\ c_2 \end{pmatrix} =$$
$$c_1 \cdot c_1^* \cdot 0 + c_1 \cdot c_2^* \cdot \tfrac{i}{2} + c_1^* \cdot c_2 \cdot \left(\tfrac{-i}{2}\right) + c_2 \cdot c_2^* \cdot 0 =$$
$$c_1 \cdot c_2^* \cdot \tfrac{i}{2} - c_1^* \cdot c_2 \cdot \tfrac{i}{2}$$
(55).

Generally, a quantum mechanical state can be described by the quantum mechanical density operator, which is given in matrix form by (*8*):

$$\rho = \overline{|\psi\rangle\langle\psi|}$$
(56),

where the overbar indicates taking an ensemble average, which means adding up the contributions from each spin in the same ensemble and then dividing by the number of spins. This matrix is calculated in similar with (12) what brings

$$\rho = \overline{|\psi\rangle\langle\psi|} = \overline{\begin{pmatrix} c_1 \\ c_2 \end{pmatrix} \cdot \begin{pmatrix} c_1^* & c_2^* \end{pmatrix}} = \begin{pmatrix} \overline{c_1 \cdot c_1^*} & \overline{c_1 \cdot c_2^*} \\ \overline{c_2 \cdot c_1^*} & \overline{c_2 \cdot c_2^*} \end{pmatrix} \quad (57).$$

The diagonal elements of the density matrix represent populations of the corresponding quantum states (*9*) and the off-diagonal elements represent coherences between corresponding quantum states. Kets and density matrices are both describing quantum-mechanical systems. The density matrix of a single particle provides exactly the same information as its ket (*10*), and therefore either is equally accurate for description of the single particle. However, the use of the density matrix is especially convenient for the description of ensembles of many spins.

The time evolution of the density operator $\hat{\rho}$ is described by the Liouville-von Neumann equation (*11*):

$$\frac{d\hat{\rho}}{dt} = -i[\hat{H}, \hat{\rho}] \quad (58).$$

When $\hat{H}$ is independent of time, the solution is given by:

$$\hat{\rho}(t) = e^{-i\hat{H}t} \hat{\rho}(0) e^{i\hat{H}t} \quad (59),$$

which is a product of three matrices, $e^{-i\hat{H}t}$, $\hat{\rho}(0)$, $e^{i\hat{H}t}$.

Then the expectation value of dynamical variable for operator $\hat{A}$ can be represented using density operator by (*12*):

$$\langle \hat{A}(t) \rangle = Tr(\hat{A} \cdot \hat{\rho}(t)) \quad (60),$$

In matrix representation *Tr* represents a sum of diagonal elements of a matrix found as a result of product of two matrices, $\hat{A}$ and $\hat{\rho}(t)$.

This description of the time evolution of the density operator, in which operators for dynamical variables remain time independent (*12*), is referred to as 'the Schrödinger representation'.

The use of density matrix instead of kets for calculation of the observables allows simplification of calculations as for kets we have to consider large number of spins and this is a many body problem that quickly becomes unmanageable due to large number of calculations required while concept of density matrix takes a different look at the problem. It does not ask how precisely each spin behaves but describes statistically what portion of spins within an ensemble is present in a particular state (diagonal elements of density matrix) and what coherences develop during spin evolution (off-diagonal elements). Such theoretical simplification allows consideration of a single matrix instead of huge number of kets in order to find the value of observable with precision which is within a very small statistical error, and therefore, is normally acceptable.

## 6. Ensemble of spins ½ in a static magnetic field

In the absence of a static magnetic field at thermal equilibrium, the spins are isotropically oriented. However, when a static magnetic field is applied (along the *z* axis, by convention) more spins in ensemble are oriented along the field than against it.
It is convenient to express the density operator in the form (*13*):

$$\hat{\rho}_0 \approx a + b\hat{I}_z \qquad (61),$$

where the constant $a = \frac{1}{2}\hat{E}$, and $\hat{E} = \begin{pmatrix} 1 & 0 \\ 0 & 1 \end{pmatrix}$. This represents the density operator in the absence of the external static magnetic field, $b\hat{I}_z$ is the part of density operator proportional to the strength of the applied static magnetic field, given by:

$$b = \frac{\gamma \hbar B_0}{ZkT} \qquad (62),$$

where $\gamma$ is magnetogyric ratio of nucleus, $\hbar$ is Planck's constant divided on $2 \cdot \pi$, $B_0$ is the magnitude of the static magnetic field, $Z$ is a partition function which is equal to 2 for spins ½ , $k$ is Boltzmann constant and $T$ is temperature (K). The ratio of the spin state populations for $|\alpha\rangle$ and $|\beta\rangle$ states is described by (*14*):

$$\frac{n_{|\alpha\rangle}}{n_{|\beta\rangle}} = e^{\frac{\Delta E}{kT}} \qquad (63),$$

where $\Delta E$ is the difference between the energies of the states.

---

**Example** Find ratio of spin state populations for proton ensemble in the static magnetic field 0.4 T at the room temperature (293K).

> *Solution:* The ratio of spin populations is given by:
> $\frac{n_{|\alpha\rangle}}{n_{|\beta\rangle}} = e^{\Delta E / kT}$, where $\Delta E$ is energy difference between $|\alpha\rangle$ and $|\beta\rangle$ states given by $\Delta E = \gamma_p \cdot \hbar \cdot |\vec{B}_0|$.
> Then $\Delta E = 267.522 \cdot 1.05458 \cdot 0.4 \cdot 10^6 \cdot 10^{-34} = 1.1285 \cdot 10^{-26} [J]$
> and $k \cdot T = 1.38 \cdot 293 \cdot 10^{-23} = 4.043 \cdot 10^{-21} [J]$.
> Then
> $\frac{\Delta E}{k \cdot T} = \frac{1.1285 \cdot 10^{-26}}{4.043 \cdot 10^{-21}} = 2.79 \cdot 10^{-6}$ and $e^{\Delta E / k \cdot T} = e^{2.79 \cdot 10^{-6}} \cong 1.00000279$.
> Hence, number of spins in $|\alpha\rangle$ state ($m = +1/2$) is 1.00000279 times larger than number of spins in $|\beta\rangle$ state ($m = -1/2$).

---

Since this ratio changes as a function of $B_0$, the net magnetization of spin ensemble builds up along the applied static magnetic field (*15*). This is called the equilibrium longitudinal spin magnetization $M_0$, which is given for an ensemble of $N$ spins 1/2 by the Curie-Weiss Law (*16*):

$$\vec{M}_0 = \frac{N\mu^2 \vec{B}_0}{kT} \qquad (64),$$

where $\mu$ is the magnitude of the magnetic moment of a spin and $T$ is the temperature.

## 7. Pulsed RF magnetic field

When a magnetic field oscillating with frequency equal to the Larmor frequency is applied perpendicular to $B_0$, its energy can be absorbed by spins. The absorbed energy changes the polarization of the spins and can transform spin states (*17*). The Larmor frequencies of nuclei in NMR experiment are in the radiofrequency (rf) range, typically tens or hundreds of MHz (or tens and hundreds millions of precession cycles per second). A rf field, applied along the $x$ axis in a Cartesian frame of reference is described by:

$$\vec{B}_{xRF} = 2B_{1x}\cos(\omega_0 t + \phi_x)\vec{x} \qquad (65).$$

The Hamiltonian of rf pulse, applied along $x$ axis in the frame of reference rotating about the $z$ axis of the laboratory frame with angular frequency $\omega_0$ in the same sense as the Larmor precession becomes time independent and can be described by:

$$\hat{H}_{xRF}^{rot} = -\gamma \hbar B_{1x} \hat{I}_x \qquad (66),$$

where only one rotating field component interacts with the spins and its energy is absorbed by spin ensemble. Propagator of this rf pulse applied for a time $t_p$ is given by:

$$U(t) = e^{-i\hat{H}_{xRF}^{rot} t_p} \qquad (67).$$

In matrix notation, which is often used in calculations, (67) can be represented as

$$U(t) = \begin{pmatrix} \cos\left(\frac{\alpha}{2}\right) & i\cdot\sin\left(\frac{\alpha}{2}\right) \\ i\cdot\sin\left(\frac{\alpha}{2}\right) & \cos\left(\frac{\alpha}{2}\right) \end{pmatrix} \qquad (68),$$

where

$$\alpha = \gamma \cdot B_1 \cdot t_p \qquad (68)$$

is an angle through which the spin magnetization rotates when an rf pulse with a frequency at the Larmor frequency of the spins and of magnitude $B_1$ is applied to the spins for a time $t_p$.

An rf pulse which rotates the spins through 90 degrees is called a 90 degree pulse; its frequency, magnitude and duration depend on the nucleus involved and on the instruments used in particular experiments.

## 8. Calculation of NMR signal

As NMR signal is proportional to transverse spin magnetization, which is detected as magnetization evolving in *xy*-plane and this can be written in the form:

$$M_{xy}(t) = Tr(\hat{\rho}(t) \cdot (\hat{I}_x + i \cdot \hat{I}_y)) \tag{69},$$

where matrix of operators $\hat{I}_x + i \cdot \hat{I}_y$ does not change with time and can be found as

$$\hat{I}_x + i \cdot \hat{I}_y = \begin{pmatrix} 0 & \frac{1}{2} \\ \frac{1}{2} & 0 \end{pmatrix} + i \cdot \begin{pmatrix} 0 & -\frac{i}{2} \\ \frac{i}{2} & 0 \end{pmatrix} =$$

$$\begin{pmatrix} 0 & \frac{1}{2} \\ \frac{1}{2} & 0 \end{pmatrix} + \begin{pmatrix} 0 & \frac{1}{2} \\ -\frac{1}{2} & 0 \end{pmatrix} = \begin{pmatrix} 0 & 1 \\ 0 & 0 \end{pmatrix} \tag{70}$$

Matrix of $\hat{\rho}(t)$ caused by 90 degree pulse applied along *x*-axis, according to (59) can be found as a product of three matrices:

$$\rho(t_{90^0}) = a + b \cdot \begin{pmatrix} \cos\left(\frac{\alpha}{2}\right) & i \cdot \sin\left(\frac{\alpha}{2}\right) \\ i \cdot \sin\left(\frac{\alpha}{2}\right) & \cos\left(\frac{\alpha}{2}\right) \end{pmatrix} \begin{pmatrix} \frac{1}{2} & 0 \\ 0 & -\frac{1}{2} \end{pmatrix} \begin{pmatrix} \cos\left(\frac{\alpha}{2}\right) & -i \cdot \sin\left(\frac{\alpha}{2}\right) \\ -i \cdot \sin\left(\frac{\alpha}{2}\right) & \cos\left(\frac{\alpha}{2}\right) \end{pmatrix} \tag{71}$$

where $\alpha = \gamma B_1 t_{90^0}$. Matrix in the middle represents $\hat{\rho}(0)$, matrices at left and right are propagator and its complex conjugate, respectively. After multiplication, the density matrix becomes:

$$\rho(t_{90^0}) = a + b \cdot \begin{pmatrix} \frac{1}{2}\cos^2\left(\frac{\alpha}{2}\right) - \frac{1}{2}\sin^2\left(\frac{\alpha}{2}\right) & -i \cdot \cos\left(\frac{\alpha}{2}\right)\sin\left(\frac{\alpha}{2}\right) \\ i \cdot \cos\left(\frac{\alpha}{2}\right)\sin\left(\frac{\alpha}{2}\right) & \frac{1}{2}\sin^2\left(\frac{\alpha}{2}\right) - \frac{1}{2}\cos^2\left(\frac{\alpha}{2}\right) \end{pmatrix} \tag{72}$$

Since $\alpha = 90^0$ the density matrix simplifies to:

$$\rho(t_{90^0}) = a + b \cdot \begin{pmatrix} 0 & -\frac{i}{2} \\ \frac{i}{2} & 0 \end{pmatrix}$$

what can be rewritten as:

$$\rho(t_{90^0}) = a + b \cdot I_y \tag{73}$$

After the rf pulse is turned off, the spins evolve in the static magnetic field $B_0$.

The Hamiltonian of this interaction is given in $\hbar$ units by:

$$\hat{H}_0 = -\gamma \cdot B_0 \cdot \hat{I}_z \tag{74}$$

Hence, density operator evolves after the rf pulse, according to:

$$\hat{\rho}(t_{90} + t_1) = e^{-i\hat{H}_0 t_1} \hat{\rho}(t_{90^0}) e^{i\hat{H}_0 t_1}$$

Then density matrix is expressed by:

$$\rho(t_{90^0} + t_1) = a + b \cdot \begin{pmatrix} e^{i\frac{\gamma \cdot B_0 \cdot t_1}{2}} & 0 \\ 0 & e^{-i\frac{\gamma \cdot B_0 \cdot t_1}{2}} \end{pmatrix} \begin{pmatrix} 0 & -\frac{i}{2} \\ \frac{i}{2} & 0 \end{pmatrix} \begin{pmatrix} e^{-i\frac{\gamma \cdot B_0 \cdot t_1}{2}} & 0 \\ 0 & e^{i\frac{\gamma \cdot B_0 \cdot t_1}{2}} \end{pmatrix} =$$

$$a + b \cdot \begin{pmatrix} 0 & -\frac{i}{2} e^{i \gamma \cdot B_0 \cdot t_1} \\ \frac{i}{2} e^{-i \gamma \cdot B_0 \cdot t_1} & 0 \end{pmatrix}$$

(75)

After decomposition of exponential:

$$e^{i \gamma \cdot B_0 \cdot t_1} = \cos(\gamma \cdot B_0 \cdot t_1) + i \cdot \sin(\gamma \cdot B_0 \cdot t_1) \tag{76},$$

it becomes:

$$\rho(t_{90^0} + t_1) =$$

$$a + b \cdot \begin{pmatrix} 0 & -\frac{i}{2}(\cos(\gamma \cdot B_0 \cdot t_1) + i \cdot \sin(\gamma \cdot B_0 \cdot t_1)) \\ \frac{i}{2}(\cos(\gamma \cdot B_0 \cdot t_1) - i \cdot \sin(\gamma \cdot B_0 \cdot t_1)) & 0 \end{pmatrix} \tag{77}.$$

This can be rewritten in the form:

$$\rho(t_{90^0} + t_1) = a + b \cdot [I_y \cdot \cos(\varepsilon(t_1)) + I_x \cdot \sin(\varepsilon(t_1))] \tag{78},$$

where $\varepsilon(t_1) = \gamma \cdot B_0 \cdot t_1$.

According to (69), transverse spin magnetization can be found as a result of the following:

$$M_{xy}(t) = Tr(\hat{\rho}(t) \cdot (\hat{I}_x + i \cdot \hat{I}_y)) =$$

$$Tr\left(\begin{pmatrix} 0 & -\frac{i}{2}(\cos(\varepsilon(t_1)) + i \cdot \sin(\varepsilon(t_1))) \\ \frac{i}{2}(\cos(\varepsilon(t_1)) - i \cdot \sin(\varepsilon(t_1))) & 0 \end{pmatrix}\begin{pmatrix} 0 & 1 \\ 0 & 0 \end{pmatrix}\right) = \tag{79}$$

$$Tr\begin{pmatrix} 0 & 0 \\ 0 & \frac{i}{2}(\cos(\varepsilon(t_1)) - i \cdot \sin(\varepsilon(t_1))) \end{pmatrix} = \frac{i}{2}(\cos(\varepsilon(t_1)) - i \cdot \sin(\varepsilon(t_1)))$$

Hence, transverse spin magnetization finally becomes:

$$M_{xy}(t) = \frac{i}{2}(\cos(\varepsilon(t_1)) - i \cdot \sin(\varepsilon(t_1))) = \frac{i}{2} e^{-i \cdot \varepsilon(t_1)} \tag{80}$$